\documentclass[journal,twoside]{IEEEtran}
\usepackage{booktabs, tabularx}
\usepackage{amsmath,amssymb,amsfonts}
\usepackage{graphicx}
\usepackage{multicol,multirow}
\usepackage{bigstrut}
\usepackage{floatrow}
\usepackage{rotating}
\usepackage{adjustbox}
\usepackage{pifont}

\def\BibTeX{{\rm B\kern-.05em{\sc i\kern-.025em b}\kern-.08em

    T\kern-.1667em\lower.7ex\hbox{E}\kern-.125emX}}

\begin{document}
\title{LDMRes-Net: Enabling Efficient Medical Image Segmentation on IoT and Edge Platforms}




\author{Shahzaib~Iqbal,
        Tariq~M.~Khan,
        Imran Razzak,
        Syed~S.~Naqvi,
        and Muhammad~Usman,
\thanks{Shahzaib and Syed S. Naqi are with the Department of Electrical and Computer Engineering, COMSATS University Islamabad (CUI), Islamabad, Pakistan.}
\thanks{Tariq M. Khan is with the School of Computer Science and Engineering, University of New South Wales, Sydney, NSW, Australia.}
\thanks{Muhammad Usman is with the Department of Computer Science and Engineering, Seoul National University, South Korea}
\thanks{Musaed Alhussein and Khursheed Aurangzeb are with the Department of Computer Engineering, College of Computer and Information, King Saud University, Saudi Arabia.}}
\maketitle

\begin{abstract}

In this study, we propose LDMRes-Net, a lightweight dual-multiscale residual block-based computational neural network tailored for medical image segmentation on IoT and edge platforms. Conventional U-Net-based models face challenges in meeting the speed and efficiency demands of real-time clinical applications, such as disease monitoring, radiation therapy, and image-guided surgery. LDMRes-Net overcomes these limitations with its remarkably low number of learnable parameters (0.072M), making it highly suitable for resource-constrained devices. The model's key innovation lies in its dual multi-residual block architecture, which enables the extraction of refined features on multiple scales, enhancing overall segmentation performance. To further optimize efficiency, the number of filters is carefully selected to prevent overlap, reduce training time, and improve computational efficiency. The study includes comprehensive evaluations, focusing on segmentation of the retinal image of vessels and hard exudates crucial for the diagnosis and treatment of ophthalmology. The results demonstrate the robustness, generalizability, and high segmentation accuracy of LDMRes-Net, positioning it as an efficient tool for accurate and rapid medical image segmentation in diverse clinical applications, particularly on IoT and edge platforms. Such advances hold significant promise for improving healthcare outcomes and enabling real-time medical image analysis in resource-limited settings.

\end{abstract}


\section{Introduction}
\IEEEPARstart{R}etinal image analysis is crucial for ophthalmologists in diagnosing blindness-causing diseases such as glaucoma, diabetic retinopathy (DR), and age-related macular degeneration (AMD), and \cite{khan2019review, iqbal2022recent, imtiaz2021screening}.  DR is one of the primary causes of preventable blindness. Franklin {\it et al.} \cite{franklin2014computerized} have stated that the characteristics of retinal blood vessels, such as slope, neovascularization, and curvature, are essential in the diagnosis of retinal diseases. Segmentation of medical images plays a crucial role in understanding image content and identifying areas of injury. It serves as the foundation for medical image analysis techniques such as medical image registration and 3D reconstruction and also has a significant impact on clinical diagnosis and treatment \cite{khan2020region}. There has been an increase in the use of deep learning as a key contribution to the field of medical image analysis. Deep learning-based medical image segmentation has become a popular research area. Many conventional semantic segmentation models, such as Fully Convolutional Networks (FCN) \cite{long2015fully} and U-Net, are based on pixel-level details \cite{zhou2019high, naveed2021towards, zhou2019unet, chen2021transunet}. 

Segmentation of medical images is a crucial task that requires the utilization of encoders and decoders. In this domain, widely recognized encoders, including VGG and ResNet, are renowned for their exceptional feature extraction capabilities \cite{gu2019net, he2016deep, valanarasu2021medical, naqvi2019automatic}. Additionally, other techniques such as multiscale grouping \cite{liu2015parsenet}, dilated convolution \cite{chen2017deeplab}, and attention mechanisms \cite{le2019classifying, ni2019raunet, li2021pyconvu} are employed to extract semantic information from medical images. Additionally, U-Net is a widely used backbone architecture for medical image segmentation due to its efficient skip connections that augment low-level features \cite{ronneberger2015u, zhou2019unet,khan2020semantically}. However, skip connections may also result in information redundancy \cite{khan2022t,khan2021residual,khan2021rc}.

\begin{figure*}
  \centering
  \includegraphics[width=1\textwidth]{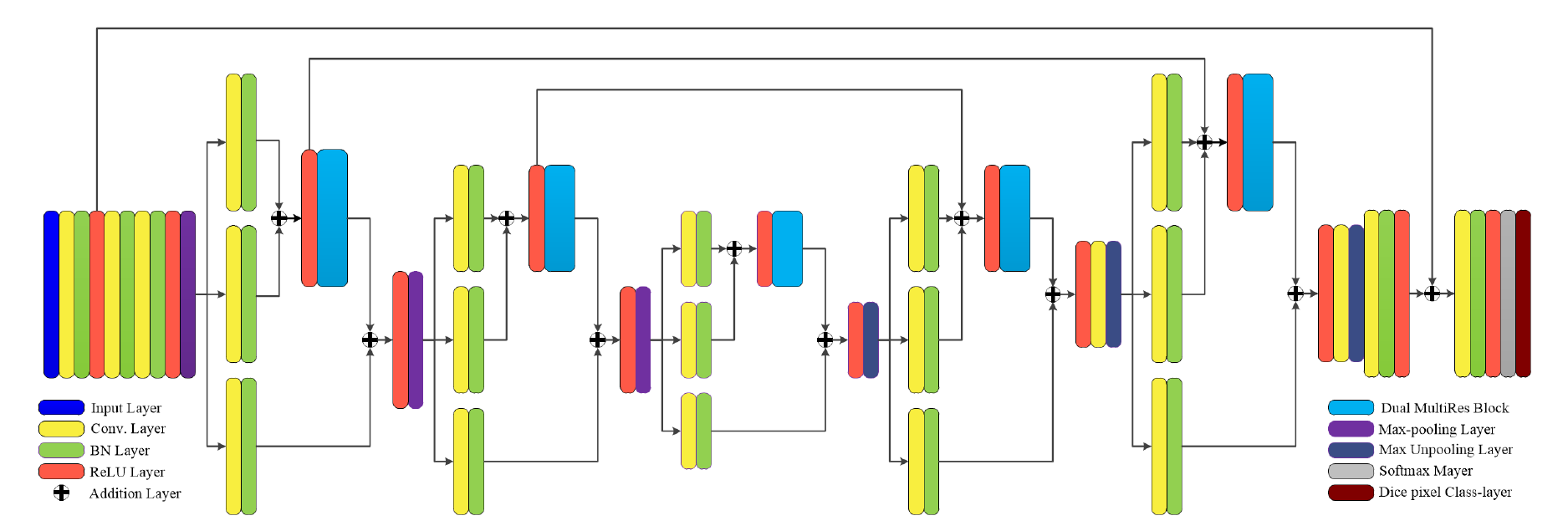}
  \caption{The illustration of the proposed LDMRes-Net architecture. The proposed architecture uses dual multi-residual blocks which is basically a combination of a multi-residual block within the residual block.}
  \label{LDMRes-Net}
\end{figure*}

In addition, the speed of segmentation is a crucial factor when using medical image segmentation models for clinical treatment. U-Net-based models are not suitable for several clinical applications, such as live disease monitoring, adaptive radiotherapy, and image-guided surgery, due to their base model complexity in comparison to segmentation performance. In practical terms, networks that are less complex and computationally efficient are required without compromising segmentation accuracy. 

In this paper, we present an innovative solution geared towards real-time systems, focusing on edge platforms and IoT devices. The proposed approach introduces the Lightweight Dual Multiscale Residual Block-based Convolutional Neural Network (LDMRes-Net), specifically tailored for retinal feature segmentation. The LDMRes-Net is a novel and lightweight dual multi-residual CNN architecture that excels in computational efficiency, boasting a mere 0.072M learnable parameters. Remarkably, this model surpasses its counterparts in accuracy without sacrificing performance. Its efficiency makes it highly suitable for resource-constrained devices, such as edge platforms and IoT devices.

At the core of LDMRes-Net lies a unique dual multiresidual block, meticulously designed to enhance both robustness and efficiency. This block capitalizes on refined feature extraction at multiple scales, significantly improving the model's ability to capture intricate details. Additionally, careful optimization of the number of filters utilized in LDMRes-Net prevents filter overlap, a common issue that could increase computational demands and training times. Consequently, the network's training time is substantially reduced, further reinforcing the practicality of LDMRes-Net for real-world applications.

In summary, the key contributions of this study are as follows.

\begin{itemize}
\item We propose a lightweight dual multiresidual CNN for retinal features segmentation, which comprises an exceptionally low number of learnable parameters (0.072M).

\item A dual multi-residual block is proposed to enhance the robustness and efficiency of the approach by leveraging the benefits of refined feature extraction and multiple scales.

\item We used the optimal number of filters to prevent filter overlap significantly reducing the training time of the proposed LDMRes-Net.

\item We tested our proposed network on multiple aspects of retinal images (such as retinal vessels, and hard exudates) to assess its robustness and generalizability.

\end{itemize}

The remaining manuscript is organized in the following way: Section \ref{sec:Related Work} provides recent relevant studies on retinal feature segmentation methodologies and lightweight medical image segmentation models. The details of the proposed model LDMRes-Net are provided in Section \ref{method}. The experimental setup is elaborated as well as the comprehensive experiments and results of LDMRes-Net are presented in Section \ref{experimentalResults}. Finally, Section \ref{sec:Conclusions} summarizes the primary findings and conclusions of the proposed work.

\begin{figure}[hb]
  \centering
  \includegraphics[width=0.85\textwidth]{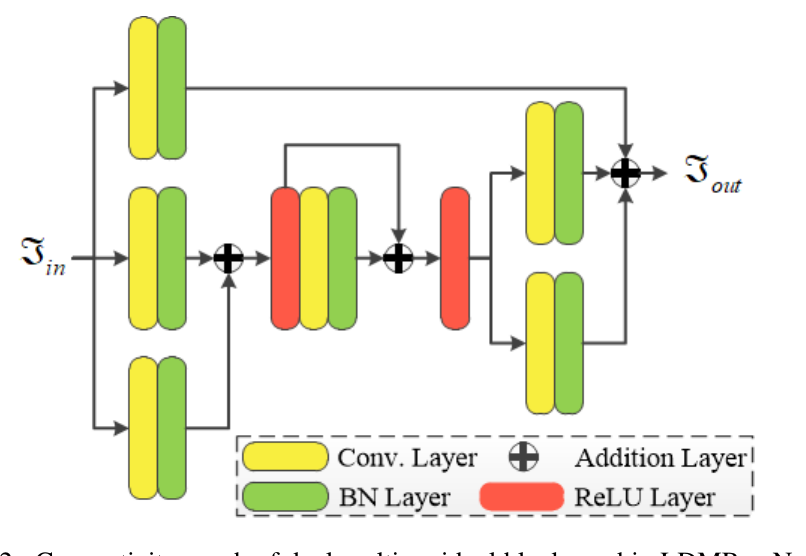}
  \caption{Connectivity graph of dual multi-residual block used in LDMRes-Net.}
  \label{multiresidual}
\end{figure}

\section{Related Work}
\label{sec:Related Work}

Medical image segmentation is an essential technique in various medical imaging applications, as it facilitates accurate diagnosis, personalized treatment planning, improved visualization, quantitative analysis, and less invasive medical procedures. U-Net architecture has undergone various adaptations and improvements in the literature, demonstrating its effectiveness for different visual analytics applications through module redesign and network construction \cite{khan2023retinal}. For instance, Milletari {\it et al.} \cite{milletari2016v} expanded U-Net to 3D image segmentation based on a fully convolutional volumetric neural network while preserving the vanilla internal structures. Contrary to the strategy used by U-Net, the method proposed by Mehta {\it et al.} \cite{mehta2017m} involves the integration of input features of varying scales at different stages. This process enables the acquisition of multilevel image information through a series of downsampling and upsampling layers. On a similar note, Zhou {\it et al.} \cite{zhou2019unet} proposed U-Net++, an approach that more efficiently captures fine-grained details through the use of densely stacked skip connections. Additionally, Oktay {\it et al.} \cite{oktay2018attention} presented Attention U-Net, which incorporates extra branches into the attention mechanism to integrate skipped and decoded data.

In recent years, lightweight segmentation networks have gained attention in image segmentation challenges. SegNAS3D \cite{wong2019segnas3d} introduced a network search to optimize the 3D segmentation structure and significantly reduce the complexity of the model. RefineNet \cite{lin2017refinenet} was used as the lightweight backbone network by Nekrasov {\it et al.} \cite{nekrasov2018light} to improve the performance of the model. IC-Net \cite{zhao2018icnet} is a real-time semantic segmentation network that uses an image cascade with branch training to accelerate model convergence. BiSeNet \cite{yu2021bisenet} is a lightweight model based on a dual-branch topology, employing multiple paths to refine spatial information. DMFNet \cite{yuan2019dmfnet} partitions channels into numerous groups and employs a weighted three-dimensional expanded convolution. This approach ultimately results in a decrease in the parameter count while simultaneously enhancing the accuracy of inference. Xception \cite{chollet2017xception} and MobileNets \cite{howard2017mobilenets, howard2019searching} use deep separable convolution to improve the inference speed. The Dense-Inception U-Net \cite{zhang2020dense} uses a compact encoder with a lightweight Inception backbone and a dense module to capture high-level semantic information. This architecture is designed for medical image segmentation tasks. 

  \begin{table*}[htbp]
  \centering
  \caption{The details of retinal fundus image datasets (publicly available) used for the performance evaluation of LDMRes-Net}
  \adjustbox{max width=\textwidth}{
  
    \begin{tabular}{lrcccccccc}
    \hline
    \multirow{3}[4]{*}{\textbf{Feature}} & \multicolumn{1}{l}{\multirow{3}[4]{*}{\textbf{Dataset}}} & \multicolumn{3}{c}{\multirow{2}[2]{*}{\textbf{Number of Images}}} & \multicolumn{1}{p{5.07em}}{\textbf{Orignal}} & \multicolumn{1}{c}{\multirow{3}[4]{*}{\textbf{FOV}}} & \multicolumn{1}{p{5.07em}}{\textbf{Resized }} & \multirow{3}[4]{*}{\textbf{Patch Size}} & \multicolumn{1}{l}{\multirow{3}[4]{*}{\textbf{Training Details}}} \\
          &       & \multicolumn{3}{c}{}  & \multicolumn{1}{p{5.07em}}{\textbf{Image }} &       & \multicolumn{1}{p{5.07em}}{\textbf{Image }} &       &  \\
\cmidrule{3-5}          &       & \textbf{Training} & \textbf{Testing} & \textbf{Total} & \multicolumn{1}{p{5.07em}}{\textbf{Resolution}} &       & \multicolumn{1}{p{5.07em}}{\textbf{Resolution}} &       &  \\
    \hline
    \multicolumn{1}{c}{\multirow{7}[2]{*}{Blood Vessels}} & \multicolumn{1}{l}{DRIVE \cite{DRIVEdata}} & 20    & 20    & 40    & $565\times 584$ & 35    & \multirow{7}[2]{*}{$640\times 640$} & \multirow{7}[2]{*}{-} & \multirow{7}[2]{*}{Image Level} \\
          & \multicolumn{1}{l}{CHASE\_DB \cite{CHASEDataset}} & 20    & -     & 20    & $960\times 990$ & 30    &       &       &  \\
          & \multicolumn{1}{l}{STARE \cite{STAREDataset}} & 28    & -     & 28    & $605\times 700$ & 45    &       &       &  \\
          & \multicolumn{1}{l}{HRF \cite{HRFDataset}} & 15    & 30    & 45    & $2336\times 3504$ & 60    &       &       &  \\
          & \multicolumn{1}{l}{ARIA \cite{ARIAdataset}} & 120   & 22    & 142   & $584\times 768$ & 50    &       &       &  \\
          & \multicolumn{1}{l}{IOSTAR \cite{IOSTARdataset}} & -     & 30    & -     & $1168\times 1752$ & -     &       &       &  \\
          & \multicolumn{1}{l}{ORVS\cite{ORVSdataset}} & 42    & 7     & 49    & $1444\times 1444$ & 45    &       &       &  \\
    \midrule
    Hard Exudates & \multicolumn{1}{l}{IDRiD \cite{IDRiDDataset}} & 54 & 27 & 81 & $2848\times 4288$ & -     & $3200\times 4480$ & $640\times 640$  & Patch Level \\
    \bottomrule
    \end{tabular}%
    }
  \label{tab: DataDescription}%
\end{table*}%

ShuffleNet \cite{zhang2018shufflenet, ma2018shufflenet} applies group convolution and channel shuffling techniques to decrease computational expenses compared to more complex models. Researchers have explored the development of lightweight segmentation networks specifically for medical images, as Iqbal {\it et al.} \cite{iqbal2022g}. However, creating networks that have low model complexity, high inference speed, and excellent performance still poses a challenge in most medical image segmentation tasks. nnU-Net \cite{isensee2021nnu} improves network adaptability by preprocessing data and postprocessing segmentation results, but the model parameters increase with this approach. U-Net++ \cite{zhou2019unet} produces high segmentation performance with minimal parameters, but its inference time is not taken into account. A lightweight V-Net \cite{lei2020lightweight} uses depth and point convolution to ensure segmentation accuracy and use fewer parameters. However, it does not speed up the inference process. Furthermore, Tarasiewicz {\it et al.} \cite{tarasiewicz2020lightweight} developed Lightweight U-Nets that can accurately delineate brain tumors from multimodal MRIs and trained several skinny networks in all image planes. PyConvU-Net \cite{li2021pyconvu} increases segmentation accuracy while using fewer parameters by replacing all traditional U-Net's convolutional layers with pyramidal convolution. However, its inference speed is low. G-Net Light \cite{iqbal2022g}, PLVS-Net \cite{arsalan2022prompt}, and MKIS-Net\cite{khan2022mkis} are effective CNN architectures for segmenting retinal blood vessels, while also being lightweight.

Transformers, unlike CNN-inspired architectures, are typically used for machine translation but have recently found applications in computer vision tasks. Vision transformers (ViTs) \cite{dosovitskiy2020image} cascade numerous transformer layers, instead of CNN-inspired architectures, and have received significant research interest. Several subsequent techniques have been developed that expand on ViTs, and many semantic segmentation algorithms \cite{zhang2022segvit, zheng2021rethinking, strudel2021segmenter, ranftl2021vision} based on ViT backbones have achieved promising results due to the powerful representations learned from pretrained backbones. However, ViTs are computationally expensive. To overcome this challenge, lightweight CNNs can be used, which offer several benefits for medical image segmentation, such as faster processing times, lower memory requirements, improved portability, lower computational cost, and lower energy consumption. Therefore, the use of lightweight methods for medical image segmentation can offer numerous advantages in terms of speed, efficiency, portability, and cost-effectiveness, making them an appealing option for a wide range of applications.

\begin{table}[htbp]
  \centering
   \caption{A performance comparison between LDMRes-Net and several alternative methods was conducted using the DRIVE \cite{DRIVEdata} dataset.}
    \adjustbox{max width=\textwidth}{%

    \begin{tabular}{lccccc}
    \toprule
    \multirow{2}[4]{*}{\textbf{Method}} & \multicolumn{5}{c}{\textbf{Performance Measures in (\%)}} \\
\cmidrule{2-6}          & \textbf{Se} & \textbf{Sp} & \textbf{Acc} & \textbf{AUC} & \textbf{F1} \\
    \midrule
    DeepVessel \cite{fu2016deepvessel} & 76.12 & 97.68 & 95.23 & 97.52 & - \\
    Orlando et al. \cite{orlando2016discriminatively} & 78.97 & 96.84 & 94.54 & 95.06 & - \\
    Att UNet \cite{oktay2018attention} & 79.46 & 97.89 & 95.64 & 97.99 & 82.32 \\
    H-DenseUNet \cite{li2018h} & 79.85 & 98.05 & 95.73 & 98.10 & 82.79 \\
    BTS-DSN \cite{Guo2019} & 78.00 & 98.06 & 95.51 & 97.96 & 82.08 \\
    DUNet \cite{Jin2019} & 79.84 & 98.03 & 95.75 & 98.11 & 82.49 \\
    M2U-Net \cite{laibacher2019m2u} & -     & -     & 96.30 & 97.14 & 80.91 \\
    Bio-Net \cite{xiang2020bio} & 82.20 & 98.04 & 96.09 & 82.06 & 98.26 \\
    CC-Net \cite{Feng2020} & 76.25 & 98.09 & 95.28 & 96.78 & - \\
    CTF-Net \cite{wang2020ctf} & 78.49 & 98.13 & 95.67 & 97.88 & 82.41 \\
    CSU-Net \cite{wang2020csu} & 80.71 & 97.82 & 95.65 & 98.01 & 82.51 \\
    OCE-Net \cite{OCE-NET} & 80.18 & 98.26 & 95.81 & 98.21 & 83.02 \\
    Wave-Net \cite{liu2022wave} & 81.64 & 97.64 & 95.61 & -     & 82.54 \\
    LightEyes \cite{guo2022lighteyes} & -     & -     & -     & 97.96 & - \\
    G-Net Light \cite{iqbal2022g} & 81.92 & 98.29 & 96.86 & -     & 82.02 \\
    \midrule

    \textbf{Proposed LDMRes-Net} & \textbf{83.58} & \textbf{98.32} & \textbf{97.02} & \textbf{98.51} & \textbf{83.09} \\
    \bottomrule
    \end{tabular}%
    }
  \label{tab: DRIVE}%
\end{table}%

\begin{figure*}[!t]
  \centering
  \includegraphics[width=0.85\textwidth]{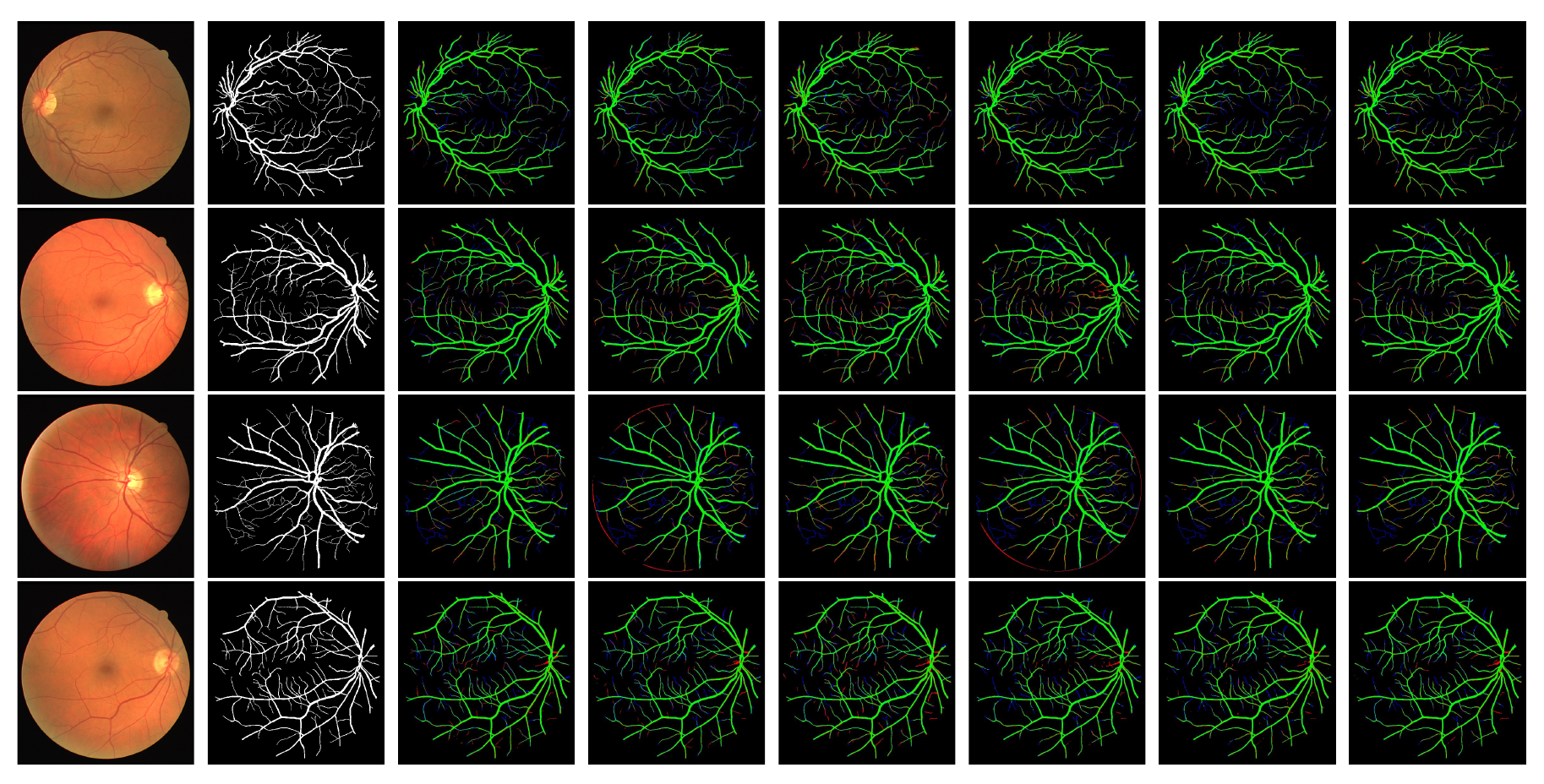}
 	\caption{Segmentation performance on several test images from the DRIVE \cite{DRIVEdata} dataset. From left to right: the input images (from top to bottom: images 1, 2, 4, 16 from the dataset), the ground truth and the results of BCD-Unet \cite{azad2019bi}, MultiResNet \cite{maji2022attention}, SegNet \cite{Badrinarayanan2017}, UNet++ \cite{zhou2018unet++}, G-Net Light \cite{iqbal2022g}, and LDMRes-Net. True positives are displayed in green, false positive pixels are displayed in red, and blue pixels represent false negatives.}
	\label{visualDRIVE}%
\end{figure*}%

\begin{table}[htbp]
  \centering
    \caption{Performance comparison of LDMRes-Net and a number of alternatives on the STARE \cite{STAREDataset} dataset.}
    \resizebox{1.0\textwidth}{!}{

    \begin{tabular}{lccccc}
    \hline
    \multirow{2}[4]{*}{\textbf{Method}} & \multicolumn{5}{c}{\textbf{Performance Measures in (\%)}} \\
\cmidrule{2-6}          & \textbf{Se} & \textbf{Sp} & \textbf{Acc} & \textbf{AUC} & \textbf{F1} \\
    \hline
    Orlando et al. \cite{orlando2016discriminatively} & 76.80 & 97.38 & 95.19 & 95.70 & - \\
    Att UNet \cite{oktay2018attention} & 77.09 & 98.48 & 96.33 & 97.00 & - \\
    BTS-DSN \cite{Guo2019} & 82.01 & 98.28 & 96.60 & 98.72 & 83.62 \\
    DUNet \cite{Jin2019} & 78.92 & 98.16 & 96.34 & 98.43 & 82.30 \\
    CC-Net \cite{Feng2020} & 80.67 & 98.16 & 96.32 & 98.33 & 81.36 \\
    OCE-Net \cite{OCE-NET} & 80.12 & 98.65 & 96.72 & 98.76 & 83.41 \\
    Wave-Net \cite{liu2022wave} & 79.02 & 98.36 & 96.41 & -     & 81.40 \\
    LightEyes \cite{guo2022lighteyes} & -     & -     & -     & 98.29 & - \\
    G-Net Light \cite{iqbal2022g} & 81.70 & 98.53 & 97.30 & -     & 81.78 \\
    \hline
    \textbf{Proposed LDMRes-Net} & \textbf{84.07} & \textbf{98.75} & \textbf{97.64} & \textbf{98.72} & \textbf{84.24} \\
    \hline
    \end{tabular}%
    }
  \label{tab:STARE}%
\end{table}%

\begin{figure*}[!t]
  \centering
  \includegraphics[width=0.85\textwidth]{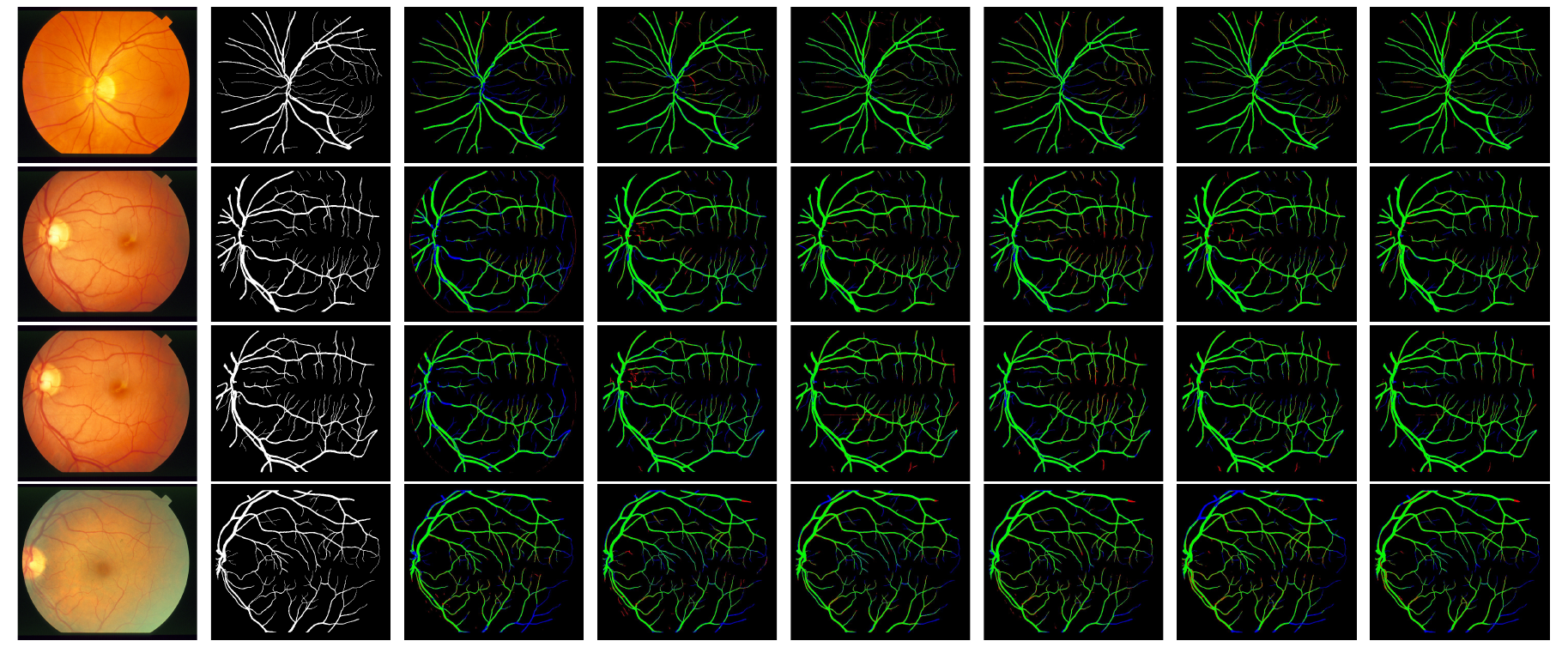}
	\caption{Segmentation performance on several test images from the STARE \cite{STAREDataset} dataset. From left side to right side: the input images (from top to bottom: images 2, 3, 4, 5 from the dataset), ground truth, and the results of BCD-Unet \cite{azad2019bi}, MultiResNet \cite{maji2022attention}, SegNet \cite{Badrinarayanan2017}, UNet++ \cite{zhou2018unet++}, G-Net Light \cite{iqbal2022g}, and LDMRes-Net. True positives are displayed in green, false positive pixels are displayed in red, and blue pixels represent false negatives.}
	\label{visualSTARE}%
\end{figure*}%

\begin{figure*}[!t]
  \centering
  \includegraphics[width=0.85\textwidth]{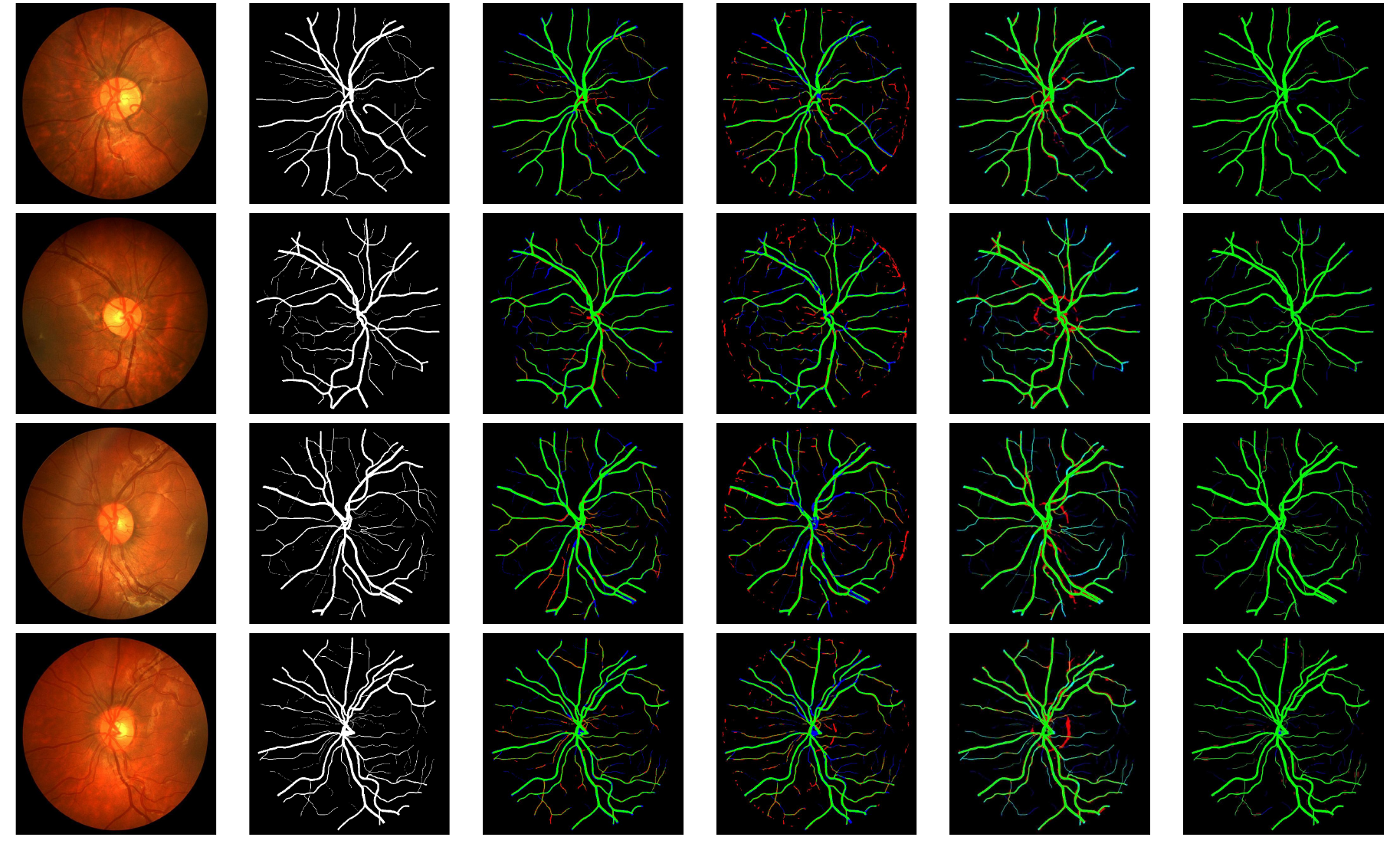}
	\caption{Segmentation results of representative test images from the CHASE\_DB \cite{CHASEDataset} dataset. From left to right: the input images, ground truth, and the results of SegNet \cite{Badrinarayanan2017}, VessSeg \cite{khan2020exploiting}, G-Net Light \cite{iqbal2022g}, and LDMRes-Net. The "red" and "blue" represents false positive and false negatives respectively.}
	\label{visualCHASE}%
\end{figure*}%

\begin{figure*}[!t]
  \centering
  \includegraphics[width=0.85\textwidth]{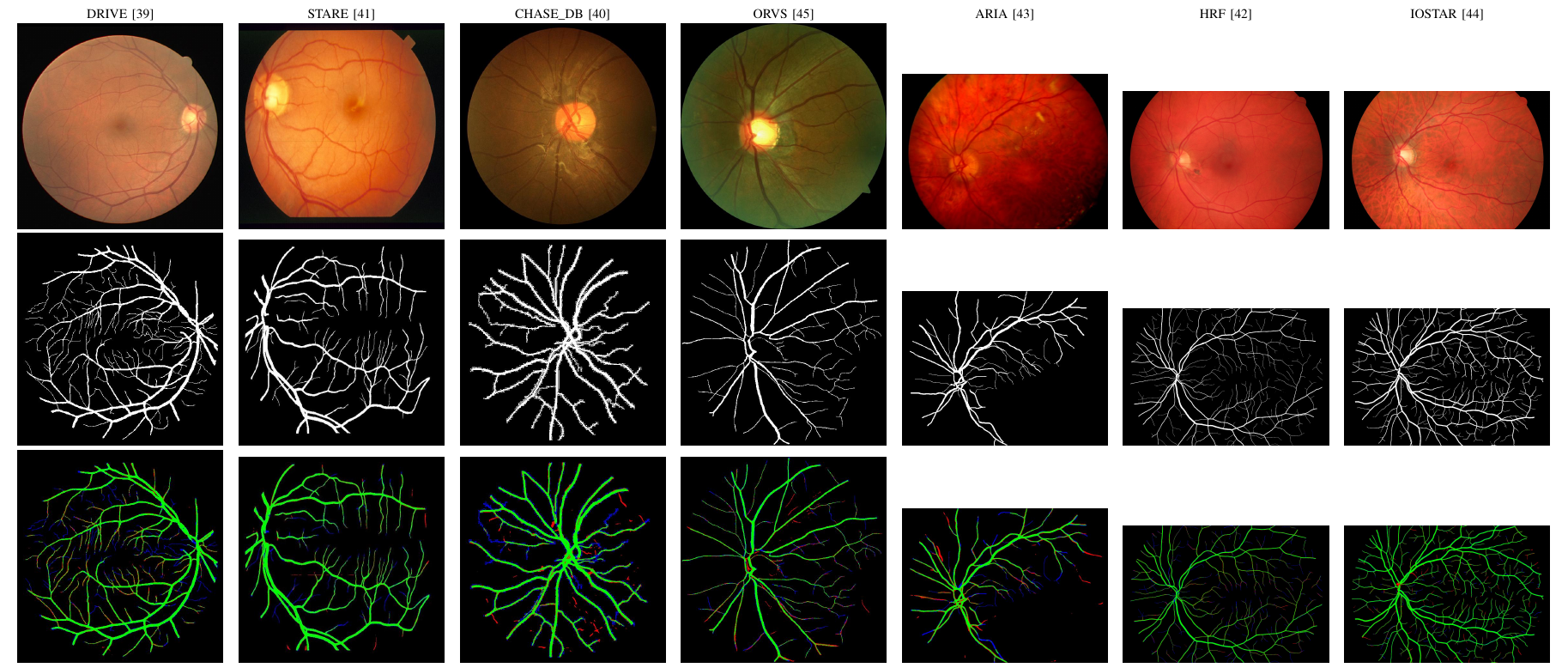}
	\caption{Visual performance of LDMRes-Net of retinal vessel segmentation on different datasets having different image resolutions.$1^{st}$ row is input RGB images, $2^{nd}$ row is corresponding ground truth images and the $3^{rd}$ row is segmentation output images of LDMRes-Net.}
	\label{fig:Otherdatasets}%
\end{figure*}%
\begin{table}[htbp]
  \centering
    \caption{A performance comparison between LDMRes-Net and several alternative methods was conducted using the CHASE\_DB \cite{CHASEDataset} dataset.}
  \resizebox{1.0\textwidth}{!}{%

    \begin{tabular}{lccccc}
    \toprule
    \multirow{2}[4]{*}{\textbf{Method}} & \multicolumn{5}{c}{\textbf{Performance Measures in (\%)}} \\
\cmidrule{2-6}          & \textbf{Se} & \textbf{Sp} & \textbf{Acc} & \textbf{AUC} & \textbf{F1} \\
    \midrule
    DeepVessel \cite{fu2016deepvessel} & 74.12 & 97.01 & 96.09 & 97.90 & - \\
    Orlando et al. \cite{orlando2016discriminatively} & 75.65 & 96.55 & 94.67 & 94.78 & - \\
    Att UNet \cite{oktay2018attention} & 80.10 & 98.04 & 96.42 & 98.40 & 80.12 \\
    BTS-DSN \cite{Guo2019} & 78.88 & 98.01 & 96.27 & 98.40 & 79.83 \\
    DUNet \cite{Jin2019} & 77.35 & 98.01 & 96.18 & 98.39 & 79.32 \\
    M2U-Net \cite{laibacher2019m2u} & -     & -     & 97.03 & 96.66 & 80.06 \\
    OCE-Net \cite{OCE-NET} & 81.38 & 98.24 & 96.78 & 98.72 & 81.96 \\
    Wave-Net \cite{liu2022wave} & 82.83 & 98.21 & 96.64 & -     & \textbf{83.49} \\
    LightEyes \cite{guo2022lighteyes} & -     & -     & -     & 98.20 & - \\
    G-Net Light \cite{iqbal2022g} & 82.10 & 98.38 & 97.26 & -     & 80.48 \\
    \midrule
    
    \textbf{Proposed LDMRes-Net} & \textbf{85.95} & \textbf{98.88} & \textbf{97.55} & \textbf{98.61} & 81.94 \\
    \bottomrule
    \end{tabular}%
    }
  \label{tab:CHASE}%
\end{table}%

\begin{table}[htbp]
  \centering
  \caption{A performance comparison between LDMRes-Net and several alternative methods was conducted using the HRF \cite{HRFDataset} dataset.}
    \resizebox{1.0\textwidth}{!}{%
  
    \begin{tabular}{lccccc}
    \toprule
    \multirow{2}[4]{*}{\textbf{Model }} & \multicolumn{5}{c}{\textbf{Performance Measures in (\%)}} \\
\cmidrule{2-6}          & \textbf{Se} & \textbf{Sp} & \textbf{Acc} & \textbf{AUC} & \textbf{F1} \\
    \midrule
    U-net \cite{ronneberger2015u} & -     & -     & 95.87 & 83.05 & 72.39 \\
    Orlando \cite{Orlando2016} & 78.74 & 95.84 & -     & -     & 95.84 \\
    DUNet \cite{Jin2019} & 74.64 & 98.74 & 96.51 & -     & - \\
    CS-Net \cite{mou2019cs} & -     & -     & 95.66 & 82.32 & 71.04 \\
    CogSeg \cite{zhang2021collaborative} & -     & -     & 96.22 & 84.31 & 74.75 \\
    SuperVessel \cite{SuperVessel} & -     & -     & 96.54 & 85.06 & 76.74 \\
    \hline
    \textbf{Proposed LDMRes-Net} & \textbf{80.78} & \textbf{98.39} & \textbf{97.11} & \textbf{89.88} & \textbf{80.20} \\
    \bottomrule
    \end{tabular}%
    }
  \label{tab:HRF}%
\end{table}%

\begin{table}[htbp]
  \centering
  \caption{Performance of LDMRes-Net ARIA \cite{ARIAdataset}, IOSTAR \cite{IOSTARdataset}, and ORVS \cite{ORVSdataset} datasets.}
  \resizebox{1.0\textwidth}{!}{%
    \begin{tabular}{lccccc}
    \toprule
    \multirow{2}[4]{*}{\textbf{Dataset}} & \multicolumn{5}{c}{\textbf{Performance Measures in (\%)}} \\
\cmidrule{2-6}          & \textbf{Se} & \textbf{Sp} & \textbf{Acc} & \textbf{AUC} & \textbf{F1} \\
    \midrule
    ARIA \cite{ARIAdataset} & 74.18 & 97.42 & 95.81 & 96.00 & 71.49 \\
    IOSTAR \cite{IOSTARdataset}  & 81.52 & 98.25 & 96.57 & 97.30 & 82.81 \\
    ORVS\cite{ORVSdataset} & 78.97 & 98.63 & 97.55 & 91.20 & 77.98 \\
    \hline
    \end{tabular}%
     }
  \label{tab:otherdata}%
\end{table}%

\section{Proposed Method}\label{method}

We propose a lightweight dual multiresidual CNN for retinal feature segmentation. The overall architecture of the proposed network is presented in (Fig.\ref{LDMRes-Net}). The proposed network is a lightweight encoder-decoder network that uses the advantages of multi-scale information extraction and feature refinement from a novel dual multi-scale residual block. The schematic connectivity of the dual multiresidual block used in LDMRes-Net is illustrated in (Fig. \ref{multiresidual}). The dual MultiRes blocks, which combine two multiscale residual blocks, offer several benefits for medical image segmentation tasks. These benefits include enhanced feature extraction, minimized overfitting, superior resolution adaptability, deeper network structure, accelerated convergence, improved handling of class imbalances, and optimized gradient flow. By utilizing two parallel multiscale residual blocks with different scale parameters, dual multiscale residual blocks can capture a wider range of features across different input image scales, resulting in diverse and complementary feature learning. Furthermore, the integration of features from different multiscale residual blocks and the use of multiple scales allows the network to generalize more effectively and reduce overfitting, particularly when dealing with limited training data. Dual multiscale residual blocks are also more resistant to fluctuations in image resolution, making the model more versatile and suitable for a broader range of segmentation tasks with varying resolutions. Additionally, the dual configuration of the multiscale residual block encourages faster convergence during training, leading to more efficient training and faster convergence to an optimal solution. Finally, the dual multiscale residual block can improve the handling of class imbalance, which is frequently encountered in retinal image feature segmentation tasks, by learning features at different scales, making the network better prepared to manage classes with different sizes and densities. The parallel arrangement of a dual multiscale residual block can also boost gradient flow during backpropagation, aiding in training deeper networks and mitigating the vanishing gradient issue. In general, dual multiscale residual blocks offer a robust and efficient approach to image segmentation tasks by taking advantage of multiple scales and refined feature extraction.

To start with, let $l^{n\times n}= \beta_{n}(f^{n \times n}(*))$ be defined as the convolution operation $f^{n \times n}$ followed by the batch normalization $beta_{n}$ on any given input image. where $n\times n$ is the size of the kernel. Let the layer input be defined as $X_{in}$, where $X\in \mathbb{R}^{C\times H\times W}$. $f_{1}$ is the output of the function in which $l^{1\times 1}$ is applied followed by an activation function calculated in (Eq. \ref{Eq1}).

\begin{equation}
    f_{1}=\Re(l^{1\times 1}(X_{in}))
    \label{Eq1}
\end{equation}

Here $\Re$ denotes the ReLU operation. Downsampled feature maps $F_{in}$, where $F_{in} \in \mathbb{R}^{C\times \frac{H}{2}\times \frac{W}{2}}$ are then computed in (Eq. \ref{Eq2}) by applying $l^{1\times 1}$ followed by $l^{3\times 3}$, an activation function, and the maximum grouping operation to the input $f_{1}$. $M_{p}$ is the max pooling operation.

\begin{equation}
    F^{enc}=M_{p}\left ( \Re\left ( l^{3\times 3}\left ( l^{1\times 1}\left ( f_{1} \right ) \right ) \right ) \right )
    \label{Eq2}
\end{equation}

Once the feature maps are down-sampled, they are fed as the input of the dual multiscale residual block, where feature extraction is performed and refined with multiscale information. The $\Im _{1}^{enc}$ is the output of the dual multiscale residual block given in (Eq. \ref{Eq:mult3}), where $k=2\times i -1$.

\begin{equation}
    \Im _{1}^{enc}=\beta _{N}(f^{1 \times 1}(F_{in}))+\sum_{i=1}^{2}\beta _{N}(f^{k\times k}(S_{2}^{enc}))
    \label{Eq:mult3}
\end{equation}

\begin{equation}
    S_{2}^{enc}=\Re\left ( \beta _{N}(f^{1 \times 1}(\breve{s}_{1})) + S_{1}^{enc} \right )
    \label{Eq:mult2}
\end{equation}

\begin{equation}
    S_{1}^{enc}=\Re\left ( \sum_{k=1}^{2}\beta _{N}\left ( f^{k\times k}\left ( l _{2} \right )) \right ) \right ) 
    \label{Eq:mult1}
\end{equation}

$S_{2}^{enc}$ and $\S_{1}^{enc}$ are the intermediate outputs of the addition layers of the dual multiscale residual block and are computed as (Eqs. \ref{Eq:mult2}-\ref{Eq:mult1}).

Once feature extraction is performed and refined at multiscale, information is fed to the ReLU layer followed by the max-pooling operation to reduce the spatial dimensions of the feature maps. The output of the second dual multiscale residual block $\Im _{2}(out)$, where $\Im _{2}(out) \in \mathbb{R}^{C\times \frac{H}{4}\times \frac{W}{4}}$ as given in (Eq. \ref{Eq1}-\ref{Eq:mult1}). In the end, the final output of the encoder $\Im _{out}^{enc}$, where $\Im _{out}^{enc} \in \mathbb{R}^{C\times \frac{H}{8}\times \frac{W}{8}}$ is computed by the ReLU layer followed by the max-pooling operation and the dual multiscale residual block. 

On the decoder side, the same process is applied to the encoder features.  The only difference on the decoder is that the max pooling $M_{p}$ layers are replaced with the upsampling layer. The output of the decoder blocks is computed as given in (Eq.\ref{Eq:mult4}-\ref{Eq:mult6}).

\begin{equation}
    \Im _{1}^{dec}=\beta _{N}(f^{1 \times 1}(\Im _{out}^{enc}))+\sum_{i=1}^{2}\beta _{N}(f^{k\times k}(S_{2}^{dec}))
    \label{Eq:mult4}
\end{equation}

\begin{equation}
    S_{2}^{dec}=\Re\left ( \beta _{N}(f^{1 \times 1}(S_{1}^{dec})) + S_{1}^{dec} \right )
    \label{Eq:mult5}
\end{equation}

\begin{equation}
     S_{1}^{dec}=\Re\left ( S_{1}^{enc} + \sum_{k=1}^{2}\beta _{N}\left ( f^{k\times k}\left ( \Im _{out}^{Enc} \right )) \right ) \right )  
    \label{Eq:mult6}
\end{equation}

\begin{equation}
    \Im _{out}^{dec}= \Gamma \left ( \Re \left ( l^{3\times 3}\left ( f_{1}+l^{3\times 3}\left ( F^{dec} \right )  \right ) \right ) \right )
    \label{Eq:decoder}
\end{equation}

The output of the decoder side $\Im _{out}^{dec} \in \mathbb{R}^{C\times H\times W}$ is computed as given in (Eq.\ref{Eq:decoder}). where, the $\Gamma$ is the softmax operation. Finally, the dice-pixel classification layer is used to obtain the binary segmentation.

\begin{table}[htbp]
  \centering
    \caption{Performance comparison of hard exudates segmentation of proposed LDMRes-Net with a number of alternatives on the IDRiD \cite{IDRiDDataset} data set.}
  \resizebox{1.0\textwidth}{!}{%
    \begin{tabular}{lcccc}
    \toprule
    \multirow{2}[4]{*}{\textbf{Method}} & \multicolumn{4}{c}{\textbf{Performance Measures in (\%)}} \\
\cmidrule{2-5}          & \textbf{Se} & \textbf{Sp} & \textbf{Acc} & \textbf{F1} \\
    \midrule
    CNN Based Model \cite{tan2017automated} & 87.58 & 98.73 & 86.32 & 85.39 \\
    Deep Residual Network \cite{mo2018exudate} & 96.30 & 93.04 & 94.08 & 95.17 \\
    Semi-Supervised Method \cite{zhou2019collaborative} & 94.38 & 94.75 & 96.73 & 95.63 \\
    L-SegNet \cite{guo2019seg} & 81.65 & 76.13 & 85.64 & 89.71 \\
    Deep Membrane System \cite{xue2019deep} & 77.91 & 96.75 & 97.31 & \textbf{96.52} \\
    CNN Based Model \cite{foo2020multi} & 85.73 & 85.62 & 90.17 & 88.93 \\
    Modified U-Net \cite{sambyal2020modified} & 93.35 & 94.82 & 95.74 & 96.81 \\
    \midrule
    \textbf{Proposed LDMRes-Net} & \textbf{96.33} & \textbf{98.98} & \textbf{98.35} & 95.05 \\
    \bottomrule
    \end{tabular}%
    }
  \label{tab:EX}%
\end{table}%


In this work, we have used the training approach using composite constraints to eliminate unnecessary filters and reduce their impact on the network learning process by promoting the sparsity proposed in \cite{khan2023neural}.

\section{Experiments and Results}
\label{sec:Results} 

In this section, we demonstrate the effectiveness of our approach for segmenting retinal blood vessels on publicly available data sets and perform a comparative analysis of our results against those achieved by previously published methods.

\subsection{Datasets}

The effectiveness of LDMRes-Net was evaluated through segmentation experiments in five different retinal image datasets (Table \ref{tab: DataDescription}). The DRIVE \cite{DRIVEdata}, STARE \cite{STAREDataset}, and CHASE\_DB \cite{CHASEDataset}, \cite{HRFDataset}, ARIA \cite{ARIAdataset}, IOSTAR \cite{IOSTARdataset}, and OVRS data sets were used to evaluate the segmentation of the retinal blood vessels and IDRiD \cite{IDRiDDataset} for the segmentation of hard exudates. Table \ref{tab: DataDescription} shows the details of the above-mentioned data sets that are used for the performance evaluation of LDMRes-Net.

\subsection{Experimental Setup}
\label{experimentalResults}

In this study, the ADAM optimizer was used with an initial learning rate set to $0.00002$ and a decay rate of $0.90$. Before the training process began, the images of each data set were standardized to dimensions of $640\times640$ pixels and then normalized using z-score statistics. This normalization process helps to ensure a more efficient learning process by keeping the input values within a similar range.

Data augmentation techniques were also incorporated to enhance the robustness of the model to various transformations. Techniques such as contrast enhancement, brightness adjustment, random flipping, and random rotation (ranging from 1 to 360 degrees) were used. These techniques artificially increased the number of training images, providing a more diverse set of data from which the model can learn.

For the specific task of segmentation of exudates, the experiments were implemented using a patch-based approach. Training images from the IDRiD dataset \cite{IDRiDDataset} were divided into non-overlapping patches of $640\times640$ pixels. Then these generated patches were partitioned into training and validation sets, adhering to a 80\%: 20\% split. This approach helps to ensure that the model is able to generalize well to unseen data and not just perform well on the data on which it was trained.

\subsection{Evaluation Criteria}
The evaluation of the segmentation results was performed by comparing them with their corresponding ground-truth images. Each pixel in the output image was classified as correctly segmented foreground pixels ({$T_{P}$}: true positives) or background pixels ({$T_{N}$}: true negatives), or as wrongly segmented foreground pixels ({$F_{P}$}: false positives) or background pixels ({$F_{N}$}: false negatives).

Sensitivity (Se) is the ratio of ({$T_{P}$}) to the total number of actual positives ({$T_{P}$+ $F_{N}$}). It represents the ability of the method to detect all positive samples correctly. It can be formulated as 

\begin{equation}
S_{e}=\frac{T_{P}}{T_{P}+F_{N}},
\label{eq:sn}
\end{equation}

Specificity (Sp) is the ratio of ({$T_{N}$) to the total number of actual negatives ({$T_{N}$+ $F_{P}$}). It represents the ability of the method to correctly identify all negative samples. It can be formulated as 

\begin{equation}
S_{p}=\frac{T_{N}}{T_{N}+F_{P}}.
\label{eq:sp}
\end{equation}

Accuracy (Acc) is the proportion of correctly identified pixels ({$T_{P}$+ $T_{N}$}) to all pixels in the image. Displays the total effectiveness of the method. It can be stated as follows. 

\begin{equation}
A_{cc}=\frac{T_{P}+T_{N}}{T_{P}+T_{N}+F_{P}+F_{N}}.
\label{eq:acc}
\end{equation}

The $F_{1}$ score, or Dice similarity coefficient (DSC), is another widely used metric to evaluate model performance. It is calculated as the harmonic mean of precision and recall, and can be expressed as:

\begin{equation}
F_{1} =\frac{2 \times T_{P}}{(2\times T_{P})+F_{P}+F_{N}}.
\label{eq:F1}
\end{equation} 

The area under the curve (AUC) is a metric used in receiver operating characteristic (ROC) analysis. The ROC curve plots the true positive rate (Se) against the false positive rate ({1+ $S_{p}$})(1-Sp) for different classification thresholds. AUC measures the overall ability of the method to distinguish between positive and negative samples. 

\begin{equation}
AUC=1-\frac{1}{2}\left(\frac{F_{P}}{F_{P}+T_{N}}+\frac{F_{N}}{F_{N}+T_{P}}\right).
\label{eq:AUC}
\end{equation}

\begin{figure*}[!t]
  \centering
  \includegraphics[width=0.85\textwidth]{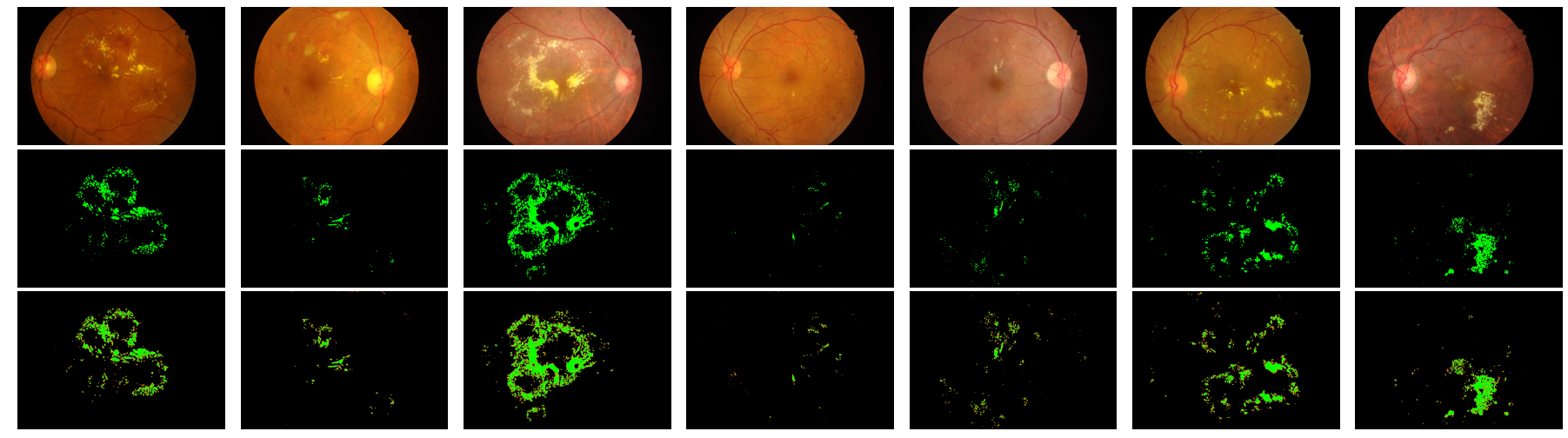}
	\caption{Hard exudates segmentation results of LDMRes-Net on representative test images from the IDRiD \cite{IDRiDDataset} dataset. $1^{st}$ row shows the RGB images, $2^{nd}$ row shows the corresponding ground truths, and $3^{rt}$ row shows the LDMRes-Net segmentation results. True positives are displayed in green, false positive pixels are displayed in red, and blue pixels represent false negatives.}
	\label{ExudatesVisuals}%
\end{figure*}%

\subsection{LDMRes-Net Generalization Analysis}

It is important to develop a method with strong generalization ability due to the variations in imaging equipment across different hospitals. The generalization ability of a model demonstrates its capacity to adapt to new data effectively. In this section, we demonstrate the generalization ability of the proposed LDMRes-Net by evaluating its segmentation performance on multiple datasets. The datasets utilized in this comparison are DRIVE \cite{DRIVEdata}, STARE \cite{STAREDataset}, CHASE\_DB1 \cite{CHASEDataset}, HRF \cite{HRFDataset}, ARIA \cite{ARIAdataset}, IOSTAR \cite{IOSTARdataset}, ORVS \cite{ORVSdataset}, and IDRiD \cite{IDRiDDataset}.

For a comprehensive understanding of the performance, we also reference prior studies that have compared numerous supervised methodologies. In particular, we assess the performance of LDMRes-Net against U-Net \cite{Olaf2015}, and SegNet \cite{Badrinarayanan2017}. These two deep-learning architectures are popular and frequently serve as benchmark models within the image segmentation community. 

Tables \ref{tab: DRIVE}, \ref{tab:CHASE}, and \ref{tab:STARE} provide the quantitative evaluation for our proposed LDMRes-Net alongside several other methodologies. As the tables illustrate, LDMRes-Net surpasses all other methods across all metrics, including sensitivity, specificity, accuracy, Area Under the Curve (AUC), and the $F1$-score.

Specifically, on the DRIVE dataset \cite{DRIVEdata}, LDMRes-Net achieved performance metrics of 83.48\%, 98.77\%, 96.97\%, 98.46\%, and 83.81\% for sensitivity, specificity, accuracy, AUC, and $F1$-score, respectively. Similarly, on the STARE dataset \cite{STAREDataset}, LDMRes-Net achieved values of 85.19\%, 98.651\%, 97.59\%, 99.06\%, and 84.22\% for sensitivity, specificity, accuracy, AUC, and $F1$-score, respectively. On the CHASE\_DB1 dataset, LDMRes-Net attained scores of 83.29\%, 98.34\%, 97.31\%, 98.82\%, and 82.10\% for sensitivity, specificity, accuracy, AUC, and $F1$-score, respectively. These findings underscore the superior performance of LDMRes-Net in comparison to other state-of-the-art (SOTA) networks. It's worth noting that among the alternative methods, there is no apparent trend or consistent superior performance in terms of specificity, AUC, or sensitivity.

We present a visual comparison of the results obtained by LDMRes-Net and alternative methods on the DRIVE \cite{DRIVEdata}, STARE \cite{STAREDataset}, and  CHASE\_DB \cite{CHASEDataset} datasets in Figures~\ref{visualDRIVE}, \ref{visualSTARE}, and \ref{visualCHASE}. The results obtained on the DRIVE dataset (Fig.\ref{visualDRIVE}) illustrate that LDMRes-Net notably reduces false positives on small vessels in comparison to current SOTA methods. U-Net-based variants, for instance, seem to struggle with accurately delineating the boundaries, while SegNet \cite{Badrinarayanan2017} tends to generate false tiny vessels across most images. The BCD-Unet \cite{azad2019bi} technique, on the other hand, seems to overlook crucial vessel information. In contrast, LDMRes-Net effectively captures this information while also minimizing the generation of false vessel information. Upon applying these methods to the STARE dataset (Fig.\ref{visualSTARE}), it's evident that alternative techniques tend to produce more false positives, especially around retinal boundaries, optic nerves, and small vessels. The proposed LDMRes-Net, however, proves to be more resilient to these artifacts in the images. Similar findings are observed when the LDMRes-Net method is applied to the CHASE\_DB dataset, as demonstrated in Fig.~\ref{visualCHASE}. This further establishes the robustness and efficacy of the LDMRes-Net method across various datasets and image conditions.

Table \ref{tab:HRF} presents a quantitative comparison of LDMRes-Net and other methods applied on the HRF dataset \cite{HRFDataset}. The LDMRes-Net outperforms other methods by achieving an average sensitivity, specificity, accuracy, AUC, and $F_{1}$-score of 80.78\%, 98.39\%, 97.11\%, 89.88\%, and 80.20\%, respectively. 
 Additionally, the performance of LDMRes-Net is evaluated on the ARIA \cite{ARIAdataset}, IOSTAR \cite{IOSTARdataset}, and ORVS \cite{ORVSdataset} datasets, as showcased in Table \ref{tab:otherdata}. Fig.\ref{fig:Otherdatasets} provides a visual representation of the vessel segmentation results obtained by our proposed LDMRes-Net on a sample image from each retinal dataset. The figure includes the corresponding ground truth and related retinal images for comparison. Despite the variations in vessel orientations and resolutions across the datasets, the figure clearly demonstrates the robust performance of LDMRes-Net in vessel segmentation tasks.

Table \ref{tab:EX} displays a quantitative comparison of the performance of LWMB-Net and other SOTA methods on the IDRiD dataset \cite{IDRiDDataset} specifically for the segmentation of hard exudates. LDMRes-Net attains an average sensitivity, specificity, accuracy, AUC, and $F1$ score of 94.35\%, 98.98\%, 98.35\%, 98.49\%, and 95.05\%, respectively. LWMB-Net outperforms other methods in every performance metric, except for the $F1$-score, where it ranks second best among the discussed methods. The Deep Membrane System \cite{xue2019deep} boasts a higher $F1$-score than the LWMB-Net, although it falls behind in other performance measures. Fig.\ref{ExudatesVisuals} visually illustrates the segmentation results for hard exudates, as achieved by LDMRes-Net on representative test images from the IDRiD dataset. Notably, LDMRes-Net attains competitive results in the presence of other lesions and features in the images, all while utilizing a minimal number of learnable parameters. This highlights the model's efficiency and effectiveness in this complex task.

\begin{figure*}[!t]
  \centering
  \includegraphics[width=0.85\textwidth]{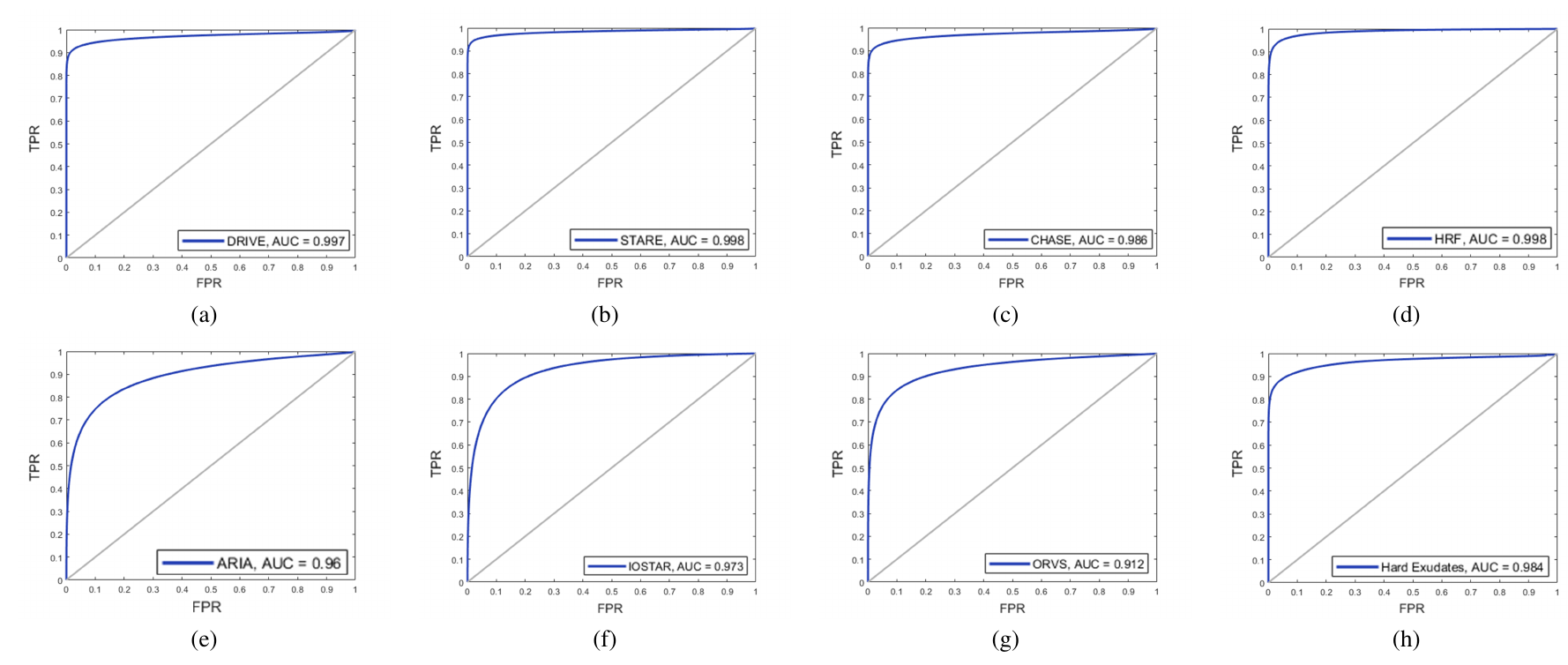}
	\caption{ROC curves of the proposed LDMRes-Net on different retinal image datasets: (a) DRIVE \cite{DRIVEdata}, (b) STARE \cite{STAREDataset}, (c)  CHASE\_DB \cite{CHASEDataset}, (d) \cite{HRFDataset}, (e) ARIA \cite{ARIAdataset}, (f) IOSTAR \cite{IOSTARdataset}, (g) ORVS\cite{ORVSdataset}, and (h) IDRiD \cite{IDRiDDataset} (hard exudates).}
	\label{fig: ROC}%
\end{figure*}%

The area under the ROC curves is a crucial metric for evaluating the performance of algorithms. In this context, to assess the impact of our proposed LDMRes-Net, we have conducted an analysis of the area under the ROC curves. Fig. \ref{fig: ROC} illustrates the ROC curves of our proposed method when applied to vessel segmentation on a variety of datasets. These datasets include DRIVE \cite{DRIVEdata}, STARE \cite{STAREDataset}, CHASE\_DB \cite{CHASEDataset}, HRF \cite{HRFDataset}, IOSTAR \cite{IOSTARdataset}, ARIA \cite{ARIAdataset}, and ORVS \cite{ORVSdataset}. The ROC curves for the segmentation of hard exudates on the IDRiD \cite{IDRiDDataset} dataset is also included. This comprehensive evaluation helps us understand the performance and reliability of the proposed LDMRes-Net across different tasks and datasets.


\section{Computational Requirements Analysis}

In this section, a thorough analysis of the computational requirements of the proposed LDMRes-Net method is presented, with a comparison to state-of-the-art (SOTA) methods. Table \ref{tab:timingComparison} is provided to offer a detailed comparison, emphasizing the efficiency and effectiveness of the LDMRes-Net approach.

\subsection{Comparison with SOTA Lightweight Networks}

Table \ref{tab:timingComparison} provides an extensive comparison between the LDMRes-Net method and various well-established state-of-the-art (SOTA) methods. The comparison primarily focuses on two crucial aspects: the number of learnable parameters and the size of the models, measured in megabytes (MB).

For comparative analysis, we include several well-established methods, namely MobileNet-V3-small \cite{Howard2019MobileNet}, ERFNet \cite{Romera2018ERFNet}, MultiRes UNet \cite{IBTEHAZ202074}, VessNet \cite{Arsalan2019}, PLVS-Net \cite{arsalan2022prompt}, M2U-Net \cite{laibacher2019m2u}, and G-Net Light \cite{iqbal2022g}. These methods have been widely recognized in the field and are commonly used as benchmarks for performance evaluation.

\subsection{Computational Efficiency of LDMRes-Net}

It should be noted that the LDMRes-Net method demonstrates remarkable computational efficiency, primarily evident in its substantially reduced number of learnable parameters. With a mere 0.072 million parameters, the LDMRes-Net surpasses other methods in terms of parameter efficiency. Importantly, this enhanced efficiency does not compromise the top-tier performance expected in medical imaging analyzes. The LDMRes-Net method achieves a harmonious balance between computational efficiency and exceptional segmentation results.

In addition, the LDMRes-Net model exhibits a remarkably compact size, requiring a mere 0.58 megabytes of storage. This compactness facilitates the effortless deployment and utilization of the LDMRes-Net method in real-world clinical settings. The reduced size of the model contributes to improved efficiency and effectiveness in medical imaging analyzes, allowing for seamless integration and practical implementation. The compact size of the LDMRes-Net model offers advantages in terms of storage requirements, enabling greater accessibility and utilization on various computing platforms.

\subsection{Practical Implications}

The computational efficiency of the LDMRes-Net method has significant practical implications. Reducing the number of learnable parameters and reducing the size of the model results in a number of advantages, including shorter training times and reduced memory requirements. Thus, LDMRes-Net is ideally adapted for environments with limited computational resources.

Moreover, the lightweight nature of LDMRes-Net enables its seamless deployment on various hardware platforms, including edge devices with restricted computational capabilities. This versatility empowers medical professionals to analyze medical images in real-time, facilitating timely decision-making and enhancing the overall efficiency of clinical workflows.

In conclusion, the complexity analysis and the comprehensive comparison presented in Table \ref{tab:timingComparison} demonstrate the computational advantages of the LDMRes-Net method. Its minimal number of learnable parameters and compact size position it as a highly efficient solution for medical image analysis. These characteristics enable its practical implementation in real-world clinical settings, contributing to more efficient and effective medical imaging analyses.


\begin{table}[!htbp]
  \centering
  \caption{Computational requirements comparisons of the LDMRes-Net with SOTA methods.}
  \resizebox{1.0\textwidth}{!}{%

    \begin{tabular}{lcc}
    \toprule
    \multicolumn{1}{c}{Method} & Params (M) & Size (MB) \\
    \midrule
    MobileNet-V3-small \cite{Howard2019MobileNet} & 2.5  & 11.0 \\
    ERFNet \cite{Romera2018ERFNet} & 2.06 & 8.0 \\
    MultiRes UNet \cite{IBTEHAZ202074} & 7.2  & - \\
    VessNet \cite{Arsalan2019} & 9.3  & 36.6  \\
    PLVS-Net \cite{arsalan2022prompt}& 1  & 3.6 \\
    M2U-Net \cite{laibacher2019m2u} & 0.55 & 2.20\\
    G-Net Light \cite{iqbal2022g} & 0.39 & 1.52 \\
    \hline
    LDMRes-Net & 0.072 & 0.58 \\
    \bottomrule
    \end{tabular}%
    }
  \label{tab:timingComparison}%
\end{table}%
\section{Conclusions}
\label{sec:Conclusions}
 This study introduces the LDMRes-Net, a retinal image feature segmentation model that exhibits exceptional performance with only 0.072 million learnable parameters. The LDMRes-Net incorporates innovative dual multiscale residual blocks, which effectively prevent feature overlap and contribute to faster convergence. The experimental results obtained with the LDMRes-Net on various retinal image features, such as retinal vessels and hard exudates, validate its robustness and versatility. Notably, the LDMRes-Net outperforms other segmentation methods in terms of retinal vessel and diabetic retinopathy (DR) lesion segmentation. The lightweight design of the LDMRes-Net positions it as an optimal choice for real-time disease diagnosis in clinical settings. With its impressive performance and computational efficiency, the LDMRes-Net holds great promise for enhancing the accuracy and efficiency of medical imaging analyses.


\end{document}